\documentclass[aps,pra,twocolumn,showpacs,10pt]{revtex4-1}
\usepackage{subfig}

\usepackage{mathptmx}
\usepackage{bbold}
\usepackage{amsmath}
\usepackage{multirow}
\usepackage{amsfonts}
\usepackage{bm}
\usepackage{amssymb}
\usepackage{hyperref}
\usepackage[pdftex]{graphicx}
\newcommand{\scs}{\scriptscriptstyle}

\begin{document}
\title{Thermo-optic bistability in 2D all-dielectric resonators}

\author{ G.V. Shadrina$^{1}$, D.N. Maksimov$^{2,3}$, and E.N. Bulgakov$^{2}$ }
\affiliation{$^1$Institute of Computational Modelling SB RAS, Krasnoyarsk, 660036, Russia}
\affiliation{$^2$Kirensky Institute of Physics, Federal Research Center KSC SB
RAS, 660036, Krasnoyarsk, Russia}
\affiliation{$^3$IRC SQC, Siberian Federal University, 660041, Krasnoyarsk, Russia}

\date{\today}
\begin{abstract}
We consider thermo-optic bistability in resonant excitation of high-quality modes in two-dimensional dielectric resonators. We develop a coupled-mode theory approach which account for the frequency shift due to a temperature dependent dielectric permittivity. The model is applied to rectangular and hexagonal resonators supporting an isolated high-quality resonant mode. The results are verified in comparison with straightforward finite-element simulations. It is shown that the model accurately describes the effect bistabily which occurs under variation of the angle of incidence or the intensity of the incident wave. In particular, it is demonstrated that variation of the incident angle can optimize the coupling between the resonator and the incident waves leading to bistabily with low intensity incident waves $W_0 = 0.35  {\rm \mu W/\mu m}^2$. The bistability threshold is shown to be extremely sensitive to the imaginary part of the dielectric permittivity $\epsilon''$.

\end{abstract}
\maketitle

\section{Introduction}

Thermophotonics is a branch of nanooptics that investigates temperature effects caused by the heating of the system by absorbed light \cite{zograf2017resonant, Khandekar17, aouassa2017temperature, Celebrano21, Cho2023, Yang2022}. Thermo-optic effects have been extensively investigated in various single-cavity setups, including plasmonic \cite{zograf2017resonant, baffou2013thermo, danesi2018photo}, all-dielectric \cite{Sun10, zograf2018local, Pottier21, Ryabov22}, and graphene-based \cite{Gao17a} structures. One of these temperature effects occurring in thermophotonic systems is nonlinearity  caused by thermorefractive phenomena \cite{jiang2020optothermal, tang2021mie, sivan2017nonlinear, li2021nonlinear, duh2020giant, zhang2020anapole}. It has been pointed out that thermo-optic effects can be dominant nonlinear effects in resonant nanophotonic structures \cite{zograf2021all}. In comparizon against plasmonic systems, all-dielectric structures \cite{kivshar2018all, baranov2017all} exhibits low material absorption, seemingly preventing effective light-to-heat conversion. However, this can be circumvented by exploiting the critical coupling effect \cite{Saadabad21, Tan22, Maksimov22}. This leads to highly efficient light absorbers \cite{zhang2015ultrasensitive, wang2020controlling, sang2021highly, xiao2021engineering, cai2022enhancing} that can be used for enhanced light-to-heat conversion.

The critical coupling effect occurs when the radiation decay rate of a resonant mode is equal to its absorption decay rate \cite{Saadabad21, Bikbaev21, Maksimov22}. Since the absorption decay rates are small in dieletrics, engineering critical coupling requires high-quality resonant eigenmodes with low radiation decay rates or, equivalently, high quality factors. There are many routs in nanophotonics that lead to high-quality resonances in dielectric structures. For example, one can employ defect modes of photonic crystals \cite{villeneuve1996microcavities,painter1999defect} or utilize optical bound states in the continuum~\cite{Hsu16, Koshelev19, sadreev2021interference}. Another approach is the application of isolated dielectric resonators, which have been shown to support high-quality resonances that can be view as quasi-bound states in the continuum \cite{rybin2017high, huang2021pushing}.

In this work we investigate resonantly enhanced thermo-optic bistability~\cite{Sun10, Khandekar17, Gao17a, Pottier21, Ryabov22,shadrina2023thermo,barulin2024thermo} in
 single dielectric resonators supporting high-quality modes. Following Huang and co-authors \cite{huang2021pushing} we will utilize high-quality modes that can be engineered in dielectric rods by constructing avoided crossings  of the
eigenvalue for pair-leaky modes. Taking into account diffusive heat transfer to the surrounding gas medium
as the cooling mechanism we will construct a non-linear model based on the temporal coupled mode theory (TCMT)~\cite{Fan03, Ruan12}
and theoretically demonstrate the effect of thermo-optic bistability due to heating by the absorbed radiation.

\begin{figure}[t]
\includegraphics[width=0.5\textwidth,height=0.16\textheight,clip]{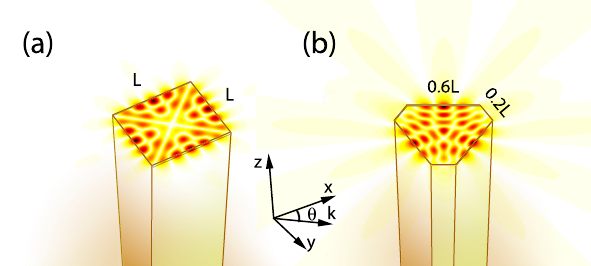}
\caption{High-quality modes supported by dielectric rods with $\epsilon=15$, $L=1 \mu m$. The modes are visualized as the absolute values of the the $z$-component of the electric field in the $x0y$-plane. (a) Dielectric rod with rectangular cross-section, $\omega_{\rm r} L/c= 6.3632-0.00018i$, $Q=1.77 \cdot 10^4$. (b) Dielectric rod with hexagonal cross-section, $\omega_{\rm r} L/c= 10.2872-0.0035i$, $Q=1.47 \cdot 10^3$. }
\label{Fig1}
\end{figure}

In this work we are going to consider the systems depicted in Fig.~\ref{Fig1}. The systems are dielectric rods of rectangular,  Fig.~\ref{Fig1}~(a), or hexagonal cross-section, Fig.~\ref{Fig1}~(b), with the real part of dielectric permittivity $\epsilon'=15$ which corresponds to dielectric rods made of silicon. The ambient medium is air $\epsilon'=1$. Assuming that the rods are infinitely extended in the $z$-direction one can analyze the problem in the framework of two-dimensional electrodynamics by considering TM-polarized waves which are described by the $z$-component of the electric field $E_z$. It is found with application of the finite-element method (FEM) together with the high-quality engineering method proposed in \cite{huang2021pushing} that the systems support isolated high-quality modes at frequencies $\omega_0 L/c=6.3632$ and $\omega_0 L/c=10.2872$, where $L=1 \mu m$ is the system size explained in Fig.~\ref{Fig1} and $c$ is the speed of light. The complex resonant eigenfrequencies $\omega_{\rm {r}}$ and the quality factors $Q$ are presented in the caption to Fig.~\ref{Fig1}. The term {\it isolated} is applied to stress that there are no other high-quality modes in the spectral vicinity of a resonance on the scale of its resonant width. In what follows we assume that the systems are illuminated by TM plane waves with the wave vector in the $x0y$-plane. The propagation direction is controlled by the angle of incidence $\theta$. We proceed to the next section for a theoretical description of scattering assuming that the dielectric permittivity is dependent on temperature.

\section{Coupled mode theory of thermo-optic bistabilty}
In this section we apply the TCMT for light scattering by an arbitrarily shaped 2D object supporting a single resonance as proposed by Ruan and Fan in \cite{Ruan12}. The starting point is
expanding the monochromatic incident and outgoing fields outside the dielectric into the Hankel functions as follows
\begin{align}\label{expansion}
& E_z^{\scs{\rm{(inc)}}}=\sum_{m=-\infty}^{\infty}a_m H_{m}^{\scs{(2)}}(k r)e^{im\phi}, \nonumber \\
& E_z^{\scs{\rm{(out)}}}=\sum_{m=-\infty}^{\infty}b_m H_{m}^{\scs{(1)}}(k r)e^{im\phi},
\end{align}
where $k$ is the vacuum wavenumber, and $r,~\phi$ are the polar coordinates. In the CGS system of units the energy fluxes in and out off the system can be calculated by the formulas below
\begin{align}\label{Flux}
    & P_{\scs{\rm{inc}}}=\frac{c}{2\pi k}\sum_{m=-\infty}^{\infty} |a_m|^2, \\
     & P_{\scs{\rm{out}}}=\frac{c}{2\pi k}\sum_{m=-\infty}^{\infty} |b_m|^2,
\end{align}
where $c$ is the speed of light in vacuum.
Since the system possesses symmetries, it is convenient to rewrite the exponentials in Eq.~\eqref{expansion} through the sine and cosine functions
as follows
\begin{align}\label{expansion2}
& E_z^{\scs{\rm{(inc)}}}=a_0^{\scs{\rm{{(c)}}}}H_{0}^{\scs{(2)}}(k r) \nonumber \\ & +\sum_{m=1}^{\infty}H_{m}^{\scs{(2)}}(k r)\left[a_m^{\scs{\rm{(c)}}}\cos(m\phi)+
a_m^{\scs{\rm{(s)}}}\sin(m\phi)\right], \nonumber \\
& E_z^{\scs{\rm{(out)}}}= b_0^{\scs{\rm{{(c)}}}}H_{0}^{\scs{(1)}}(k r)\nonumber \\
& + \sum_{m=1}^{\infty}H_{m}^{\scs{(1)}}(k_0 r)\left[b_m^{\scs{\rm{(c)}}}\cos(m\phi)+
b_m^{\scs{\rm{(s)}}}\sin(m\phi)\right].
\end{align}
Equation~\eqref{Flux} can be rewritten in terms of the new expansion coefficients 
\begin{align}\label{Flux2}
    & P_{\scs{\rm{inc}}}=\frac{c}{2\pi k}\left[|a_0^{\scs{\rm{(c)}}}|^2+\frac{1}{2} \sum_{m=1}^{\infty} (|a_m^{\scs{\rm{(c)}}}|^2+|a_m^{\scs{\rm{(s)}}}|^2)\right], \\
     & P_{\scs{\rm{out}}}=\frac{c}{2\pi k}\left[|b_0^{\scs{\rm{(c)}}}|^2+\frac{1}{2} \sum_{m=1}^{\infty} (|b_m^{\scs{\rm{(c)}}}|^2+|b_m^{\scs{\rm{(s)}}}|^2)\right].
\end{align}
In this work we assume that the system is illuminated by a plane wave. For applying Eq.~\eqref{expansion2} the incident plane wave is expanded into the Bessel function as follows
\begin{equation}
    e^{i{\bf k}{\bf r}}=\sum_{m=-\infty}^{\infty}i^me^{-i\theta}J_m(kr)e^{im \phi},
\end{equation}
where $\theta$ is the angle of incidence, and $\bf{k}$ is the vacuum wave vector.
Note that the Bessel function of the first kind contains both outgoing and incoming Hankel functions 
\begin{equation}
    J_m(kr)=\frac{1}{2}\left[H_{m}^{\scs{(1)}}(k r)+H_{m}^{\scs{(2)}}(k r) \right].
\end{equation}
Therefore, for a correct application of the TCMT, we have to write the incident field as follows
\begin{align}\label{exp_final}
   & E_z^{\scs{\rm{(inc)}}}=\frac{1}{2}H_{0}^{\scs{(2)}}(k r)  +\sum_{m=1}^{\infty}i^{m}H_{m}^{\scs{(2)}}(k r)\cos[m(\phi-\theta)]. 
\end{align}

According to \cite{Ruan12} the TCMT solution for the scattering matrix takes the following form
\begin{equation}
    \widehat{S}=\frac{{\bf d}{\bf \kappa}^{\intercal}}{i(\omega_0-\omega)+\gamma},
\end{equation}
where $\omega$ is the incident frequency, $\omega_0$ is the center-frequency of the resonant mode,  $\gamma$ is the width of the resonance, and ${\bf \kappa},~{\bf d}$ are the coupling and the decoupling vectors, correspondingly. The coupling vector is assembled of coupling constants between the resonant mode and the incident channels. The decoupling vector ${\bf d}$ can be related to ${\bf \kappa}$ as demonstrated in \cite{Ruan12}. This, however, is not needed for our purpose, since here we are only
interested in heating the resonator by incident light. Therefore, for our purpose it suffices to use the TCMT solution for the squared absolute value of the resonant mode amplitude
\begin{equation}\label{alinear}
    |a|^2=\frac{|{\bf \kappa}^{\intercal}{\bf s}_{\scs{\rm in}}|^2}{(\omega_0-\omega)^2+\gamma^2},
\end{equation}
where ${\bf s}_{\scs{\rm in}}$ is the vector of the expansion coefficients of the plane wave into the incident channel functions. The TCMT solution for the scattering field within the dielectric is, thus, written as
\begin{equation}\label{field}
    E_z(r,\phi)=aE_z^{\scs{(0)}}(r,\phi),
\end{equation}
where $E_z^{\scs{(0)}}(r,\phi)$ is the eigenfield of the resonant mode.

The key ingredients to correct application of the TCMT are normalization conditions for both the resonant eigenmode and the channel functions. The resonant eigenmode has to be normalized to store a unit energy in the scattering domain 
\begin{equation}\label{norm}
    \frac{\epsilon'}{8\pi}\int|E_z^{\scs{\rm (0)}}|^2dS=1.
\end{equation}
 At the same time the incident channel functions have to be normalized to supply a unit flux into the scattering domain. Therefore, the incident Hankel functions in Eq.~\eqref{exp_final} have to re-normalized in accordance with Eq.~\eqref{Flux2}. For a plane wave $e^{i{\bf k}{\bf r}}$ the elements of 
\begin{equation}\label{s_in}
{\bf s}_{\scs{\rm in}}=(s_0,~s_1^{\scs{\rm (c)}},~s_1^{\scs{\rm (s)}},~\ldots,~s_m^{\scs{\rm (c)}},~s_m^{\scs{\rm (s)}},~\ldots)
\end{equation}
are as follows
\begin{align}
    & s_0=\sqrt{\frac{c}{8\pi k}}E_0, \nonumber \\
    &  s_{m}^{\scs{\rm (c)}}=\sqrt{\frac{c}{4\pi k}}i^m\cos(m\theta)E_0, \nonumber \\
    & s_{m}^{\scs{\rm (s)}}=\sqrt{\frac{c}{4\pi k}}i^m\sin(m\theta)E_0,
\end{align}
where $E_0$ is a coefficient in front
of $e^{i{\bf k}{\bf r}}$ that is introduced to control the intensity of the incident wave with $E_0=0$ corresponding to a unit flux.
For consistency with Eq.~\eqref{alinear} the elements of the coupling vector are in what follows enumerated in the same manner as in Eq.~\eqref{s_in}
\begin{equation}\label{kappa}
{\kappa}=(\varkappa_0,~\varkappa_1^{\scs{\rm (c)}},~\varkappa_1^{\scs{\rm (s)}},~\ldots,~\varkappa_m^{\scs{\rm (c)}},~\varkappa_m^{\scs{\rm (s)}},~\ldots)
\end{equation}
It is demonstrated in~\cite{Ruan12} that the coupling vector ${\bf \kappa}$ can be found numerically by expanding the resonant mode 
into the outgoing cylindrical harmonics in the far field, see the second line of Eq.~\eqref{expansion2}. One should remember, however,
that in applying Eq.~\eqref{expansion2} on has to use
the complex wave number
\begin{equation}\label{complex}
    \bar{k}=\frac{1}{c}\left(\omega_0-i{\gamma}\right)
\end{equation}
corresponding the complex eigenfrequency of the resonant mode.
The calculated vector of the expansion coefficients of the resonant eigenmode ${\bf b}_{\rm r}$ has to be normalized to respect the energy conservation condition~\cite{Ruan12}
\begin{equation}\label{normgamma}
    2\gamma_{\rm{r}}=\kappa^{\dagger}\kappa,
\end{equation}
where $\gamma_{\rm{r}}$ is the radiative decay rate. The imaginary part of the complex eigenvalue, Eq.~\eqref{complex} is given by \cite{christopoulos2019calculation}
\begin{equation}
    \gamma=\gamma_{\rm{r}}+\gamma_a,
\end{equation}
where $\gamma_a$ is the absorption decay rate. The absorption decay rate, in its turn, can be calculated by integrating the scattering solution over the area of the resonator and taking into account that the absorbed power is given by
\begin{equation}\label{Pa}
     P_a=\frac{\omega\epsilon''}{8\pi}\int|E_z|^2dS,
\end{equation}
where $\epsilon''$ is the imaginary part of the dielectric permittivity.
After applying the normalization condition Eq.~\eqref{norm} one finds 
\begin{equation}
   {\gamma_{a}}=\frac{\omega \epsilon''}{2\epsilon'}.
\end{equation}
Now it is possible to calculate the radiation decay rate $\gamma_{\rm{r}}$ and
write down the coupling vector that complies with Eq.~\eqref{normgamma} as follows
\begin{equation}
    \kappa={\bf b}_{\rm r}\frac{\sqrt{2\gamma_{\rm{r}}}}{{\bf b}_{\rm r}^{\dagger}{\bf b}_{\rm r}}.
\end{equation}

Now our goal is incorporating thermo-optic effects into the TCMT model. We start from a linear dependence of the refractive index on temperature
\begin{equation}
    n=n+n_1\Delta T, \ n_1=2\times10^{-4} \frac{1}{K}.
\end{equation}
The increment of temperature is proportional to  the absorbed radiation power in the bulk dielectric
\begin{equation}
    \Delta T=\bar{\varkappa}P_a,
\end{equation}
which is a consequence of linearity of the heat equation. The coefficient $\bar{\varkappa}$ is to be calculated by numerically solving
the heat equation with a zero temperature increment at a distance from the microresonator.
The quantity $P_a$ can be found from Eq.~\eqref{Pa}. After applying Eq.~\eqref{field} together with the normalization condition Eq.~\eqref{norm} we find for the increment of the real part of the dielectric permittivity \begin{figure}[t]
\includegraphics[width=0.4\textwidth,height=0.2\textheight, trim = 0cm 6.5cm 0cm 6.5cm, clip]{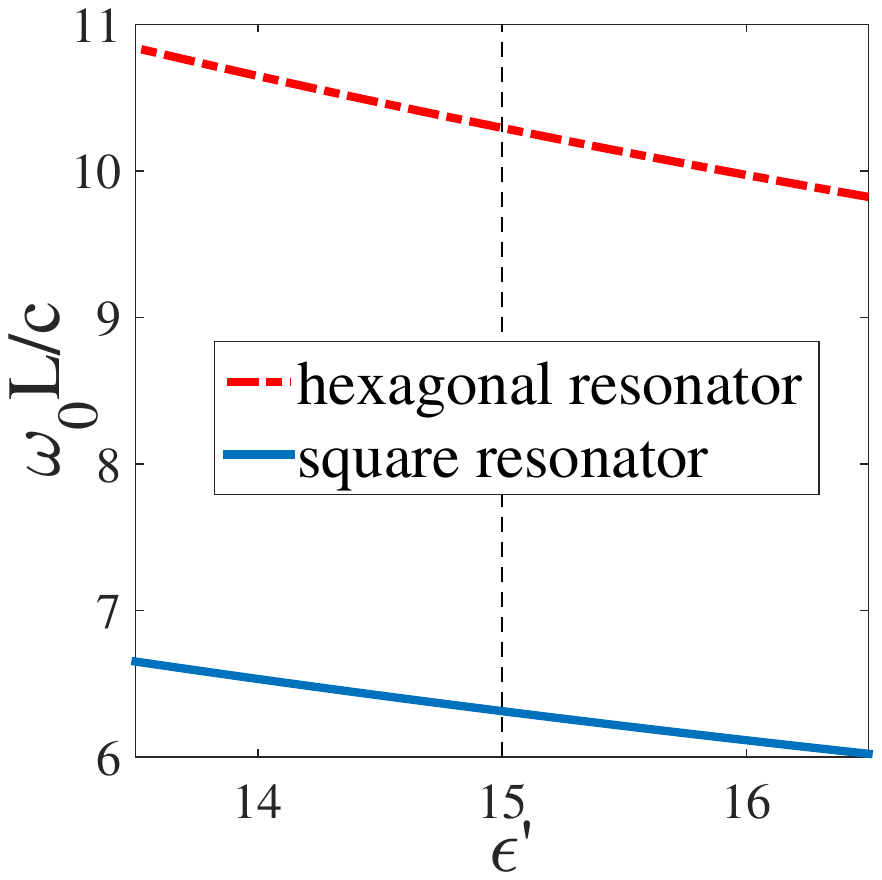}
\caption{Resonant eigenfrequency $\omega_0$ against the real part of the dielectric permittivity $\epsilon'$. In the vicinity of $\epsilon'=15$; square resonator, $\Delta\omega_0=-0.491\omega_0\Delta\epsilon'/\epsilon'$; hexagonal resonator, $\Delta\omega_0=-0.495\omega_0\Delta\epsilon'/\epsilon'$.}
\label{Fig1_5}
\end{figure}
\begin{equation}\label{deps}
    \Delta\epsilon'=2\frac{n_1\epsilon''\bar{\varkappa}}{\sqrt{\epsilon'}}\omega|a|^2.
\end{equation}
\begin{figure*}[th]
\includegraphics[width=1\textwidth,height=0.40\textheight,clip]{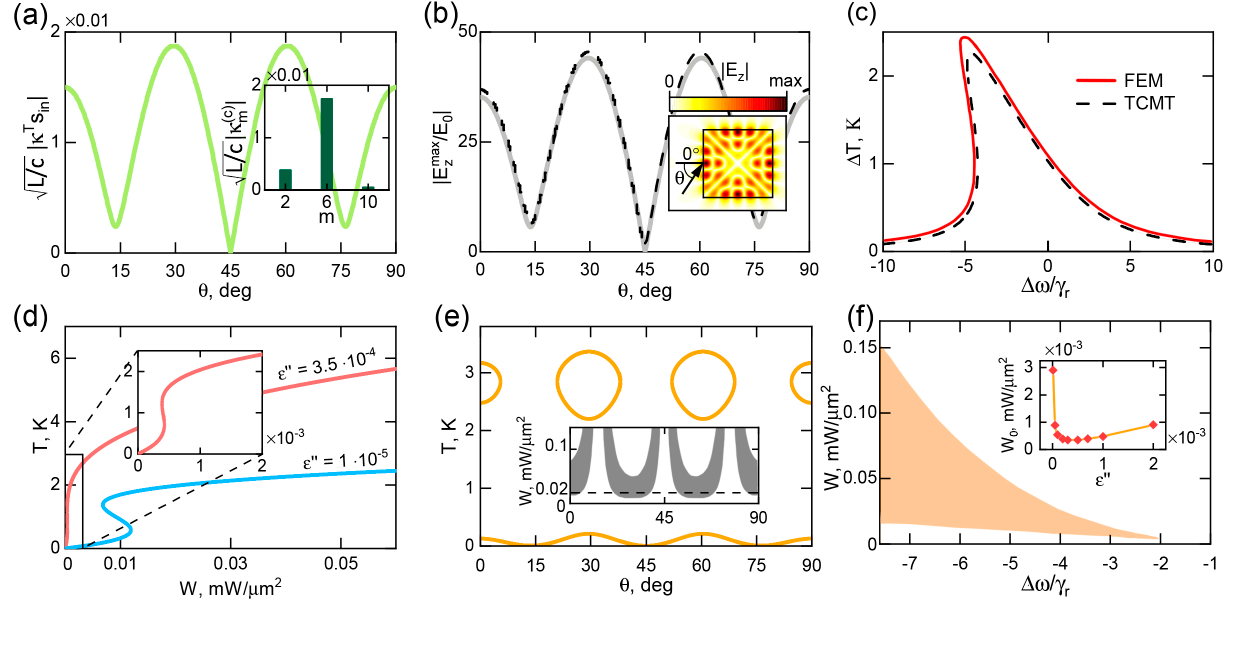}
\caption{Thermo-optic bistability in a square resonator. 
(a) The absolute value of the scalar product between the coupling vector and the incident field $\sqrt{L/c}|{\bf \kappa}^{\intercal}{\bf s}_{\scs{\rm in}}|$ as a function of the incident angle. The inset shows the absolute values of the largest elements of the coupling vector. 
(b) Resonant field enhancement against the angle of incidence with $E_z^{\rm{max}}$ as the maximal value of the electric field in the resonator. The mode is excited by a plane wave. The solid line shows the prediction of the TCMT model. The dashed line shows the results of direct simulations. The incident frequency is $\omega L/c=6.3632$. The inset shows the field profile at the incident angle $\theta=0$. 
(c) The temperature increment against the frequency detuning $\Delta\omega=\omega-\omega_0$ at the incident angle $\theta=0$ and intensity $W=5.7 \cdot 10^{-4} \rm{mW/\mu m^2}$, $\epsilon''=7 \cdot 10^{-4}$ in the critical coupling regime $\gamma_{\rm r} = \gamma_a$. The solid line shows the results of direct simulations. The dashed line shows the prediction of the TCMT model. 
(d) Thermo-optic bistability in the intensity domain for two different values of the imaginary part of the dielectric permittivity $\epsilon''$ at $\Delta\omega/{\gamma_{\bf r}}=-3$. (e) The temperature increment as a function of the angle of incidence at $\Delta\omega/{\gamma_{\rm r}}=-6$ and the incident intensity $W=0.02 \rm{mW/\mu m^2}$, $\epsilon''=1 \cdot 10^{-5}$. The inset shows the bistability domains as shaded areas in space of the angle of incidence and intensity. 
(f) The shaded area shows the bistability domain in space of frequency detuning and the intensity of the incident waves, $\theta=0$, $\epsilon''=1 \cdot 10^{-5}$. The inset shows the critical value of intensity at which the bistability occurs as a function of the imaginary part of the dielectric permittivity, $\theta=0$.}
\label{Fig2}
\end{figure*}
The frequency shift due to heating can be assessed from Maxwell's equations for a closed resonator as follows
\begin{equation}\label{freq}
    \Delta\omega_0=-\frac{1}{2}\frac{\omega_0}{\epsilon'}\Delta\epsilon'.
\end{equation}
The above equation is confirmed by direct numerical simulations, see the data in Fig.~\ref{Fig1_5}.
After combining Eq.~\eqref{freq} and Eq.~\eqref{deps} we have
\begin{align}
& \Delta\omega_0=-\chi\omega|a|^2, \nonumber \\
& \chi=\frac{n_1\epsilon''\bar{\varkappa}\omega_0}{\epsilon'\sqrt{\epsilon'}}.    
\end{align}
Finally, in accordance with Eq.~\eqref{alinear} the non-linear TCMT equation that accounts for change of the real part of the dielectric permittivity is written as follows
\begin{equation}\label{final}
    |a|^2=\frac{|{\bf \kappa}^{\intercal}{\bf s}_{\scs{\rm in}}|^2}{[\omega(1+\chi|a|^2)-\omega_0]^2+\gamma^2},
\end{equation}
After solving Eq.~\eqref{final} the temperature increment can be found as
\begin{equation}
    \Delta T=\frac{\epsilon''\bar{\varkappa}}{\epsilon'}\omega|a|^2.
\end{equation}

\section{Numerical simulations}
In this section we apply the TCMT approach for describing the effect of thermo-optic bistability induced by excitation of the resonant modes shown in Fig.\ref{Fig1}. In both cases the TCMT results shall be compared with the direct numerical solution of the scattering problem by the FEM. The heat equation is also solved with the FEM
with the thermal conductivities 
\begin{align}
& \varkappa= 156 \  \frac{W}{m\cdot K} \ \ {\rm in \ silicon}, \nonumber \\
& \varkappa= 0.02 \  \frac{W}{m\cdot K} \ \ {\rm in \ air}.
\end{align}

\begin{figure*}[th]
\includegraphics[width=1\textwidth,height=0.35\textheight,clip]{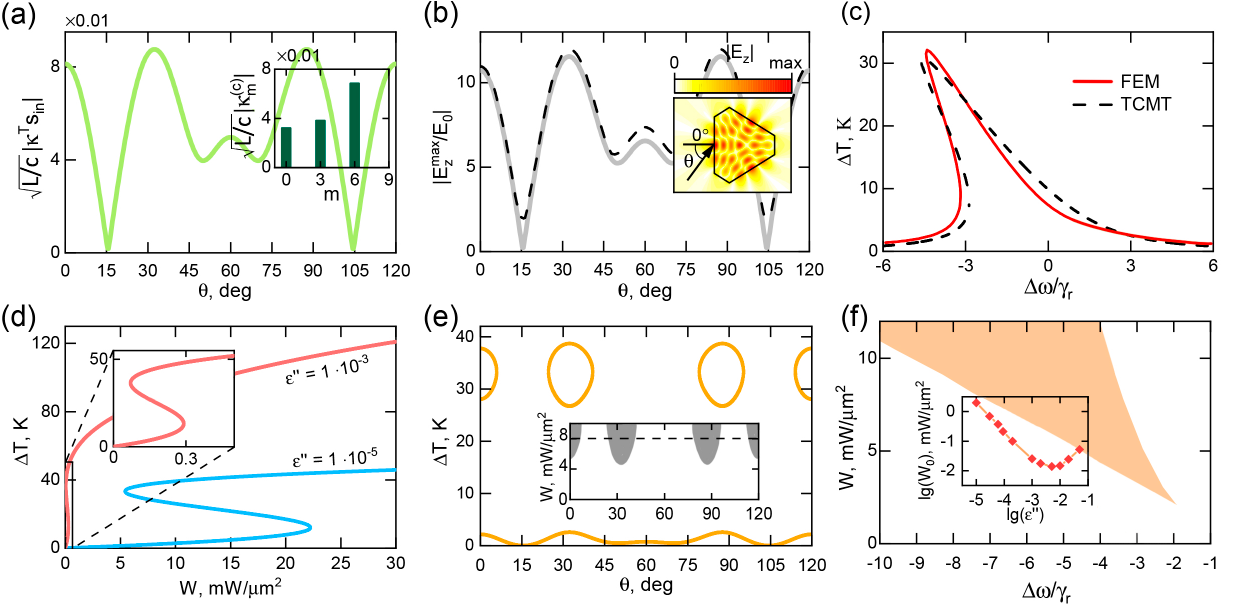}
\caption{Thermo-optic bistability in a hexagonal resonator. 
(a) The absolute value of the scalar product between the coupling vector and the incident field $\sqrt{L/c}|{\bf \kappa}^{\intercal}{\bf s}_{\scs{\rm in}}|$ as a function of the incident angle. The inset shows the absolute values of the largest elements of the coupling vector. 
(b) Resonant field enhancement against the angle of incidence with $E_z^{\rm{max}}$ as the maximal value of the electric field in the resonator. The solid line shows the prediction of the TCMT model. The dashed line shows  the results of direct simulations. The incident frequency is $\omega L/c = 10.2872$. The inset shows the field profile at the incident angle $\theta=0$.
(c) The temperature increment against the frequency detuning $\Delta\omega=\omega-\omega_0$ at the incident angle $\theta=0$ and intensity $W=5\rm{mW/\mu m^2}$, $\epsilon''= 1 \cdot 10^{-5}$. The solid line shows the results of direct simulations. The dashed line shows the prediction of the TCMT model. 
(d) Thermo-optic bistability in the intensity domain for two different values of the imaginary part of the dielectric permittivity $\epsilon''$ at $\Delta\omega/\gamma_{\rm r}=-5$.
(e) The temperature increment as function of the angle of incidence at $\Delta\omega/\gamma_{\rm r}=-5$ and the incident intensity $W=8 \rm{mW/\mu m^2}$, $\epsilon''=1 \cdot 10^{-5}$. The inset shows the bistability domains as shaded areas in space of the angle of incidence and intensity. 
(f) The shaded area shows the bistability domain in space of frequency detuning and the intensity of the incident waves, $\theta=0$, $\epsilon''=1 \cdot 10^{-5}$. The inset shows the critical value of intensity at which the bistability occurs as a function of the imaginary part of the dielectric permittivity at $\theta=0$.}
\label{Fig3}
\end{figure*}

The high-quality mode in the rectangular resonator shown in Fig.~\ref{Fig1}(a) is symmetric with respect to both $x\rightarrow -x$ and $y\rightarrow -y$, but anti-symmetric with respect to inversion of the diagonal axes of the square.  Due to this symmetry we have 
the only non-zero coupling constants
\begin{equation}
    \varkappa_{\ell}^{\scs{\rm (c)}}\neq 0, \ \mathrm{if} \ \ell=2,6,10,...
\end{equation}
In Fig.~\ref{Fig2}~(a) we show the angular dependence of the dimensionless quantity $\sqrt{L/C}|{\bf \kappa}^{\intercal}{\bf s}_{\scs{\rm in}}|$ which occurs in the numerator of Eq.~\eqref{final}. The absolute values of the three larges coupling coefficients are shown in the inset in the same subplot. One can see in Fig.~\ref{Fig2}~(a) that the efficiency of coupling is strongly dependent on the angle of incidence. This observation is confirmed in Fig.~\ref{Fig2}~(b) where we compare the TCMT and the FEM solutions of the optical problem in the linear regime. The strength of the resonant excitation is characterized by the ratio of the maximal value of the electric field in the resonator $E_z^{\rm \max}$ to the amplitude of the incident field $E_0$. One can see in Fig.~\ref{Fig2}~(b) that the TCMT solution closely follows the result of full-wave simulation with the FEM. Furthermore, as one can see in the inset in Fig.~\ref{Fig2}~(b), the field profile of the scattering solution is dominated by the resonant mode shown in Fig.~\ref{Fig1}~(a). The TCMT and the FEM solutions of the full problem with account of the temperature dependence of the refractive index are shown in Fig.~\ref{Fig2}~(c) where one can see a typical picture of the non-linear resonance in the frequency domain. As before we observe a good coincidence between between the TCMT and the full-wave simulations. In Fig.~\ref{Fig2}~(d) we demonstrate the effect of bistability in the intensity domain. In full accordance with Eq.~\eqref{deps} one observes that the bistability threshold is dependent on the imaginary part of the dielectric permittivity. The sensitivity of the coupling efficiency shown in Fig.~\ref{Fig2}~(a) suggests that the bistability can be controlled by variation of the angle of incidence. The suggestion is confirmed in Fig.~\ref{Fig2}~(e) where we show the solutions of Eq.~\eqref{final} for different values of $\theta$. On can see that with the incident intensity $W=0.02 {\rm mW/\mu m}^2$ the bistability occurs only at certain angles of incidence which correspond to the maxima of the resonant field enhancement in Fig.~\ref{Fig2}~(b). In the inset in Fig.~\ref{Fig2}~(e) we show the bistability domain in the space of incident intensity and angle. One can see in the inset that the bistability threshold is lowered significantly at the angle of maximal resonant field enhancement. Finally, in Fig.~\ref{Fig2}~(f) we demonstrate the bistability domain in space of the incident frequency and the incident intensity at $\theta=0$. One can see that with $\epsilon''=1 \cdot 10^{-5}$ the bistability threshold is of the value $W_0\approx 0.01  {\rm mW/\mu m}^2$. Besides tuning the incident frequency to the peak of the non-linear resonance resulting to the maximal field enhancement, the bistability is also controlled by the efficiency of absorption. To examine this effect we ran numerical simulation for different values of $\epsilon''= 1 \cdot 10^{-5} - 2 \cdot 10^{-3} $. The results are shown in the inset in Fig.~\ref{Fig2}~(f). One can see in the inset that the threshold value of the incident intensity resulting in the bistability reaches minimum at a certain value of $\epsilon''=5 \cdot 10^{-4}$ that corresponds to the critical coupling, $\gamma_{a}=\gamma_{\rm r}$. In fact the frequency domain response in the near to critical coupling regime is already demonstrated in Fig.~\ref{Fig2}~(c), hence the low value of the incident intensity $W=5.7 \cdot 10^{-4} \rm{mW/\mu m^2}$. According to the data presented in the inset to Fig.~\ref{Fig2}~(f) the lowest threshold intensity that can be achieved with the square resonator in the critical coupling regime is $W_0 = 0.35  {\rm \mu W/\mu m}^2$. The temperature difference between the two stable solutions is, though, vanishingly small.

In the case of the hexagonal resonator the resonant mode is symmetric with respect to all symmetry axes leading to the non-zero coupling coefficients
\begin{equation}
    \varkappa_{\ell}^{\scs{\rm (c)}}\neq 0, \ \ \ell=0,3,6,...
\end{equation}
In Fig.~\ref{Fig3} we present the same set of data for the hexagonal resonator as shown of the rectangular resonator in Fig.~\ref{Fig2}. One can see in Fig.~\ref{Fig3} that all observations made for the rectangular resonator hold with exception that the bistability threshold is now obtained with two orders of magnitude higher incident intensities. This is due to the lower quality factor of the eigenmode, see the caption to Fig.~\ref{Fig1}. 
\section{Summary and Conclusions}

We demonstrated thermo-optic bistability in resonant excitation of high-quality modes in two-dimensional dielectric resonator. We proposed a coupled-mode theory approach which account for the frequency shift due to a temperature dependent dielectric permittivity. The approach only requires calculating the complex eigenfrequency and the eigenfield of the resonant eigenmode which are used for calculating all parameters in a single non-linear equation, Eq.~\eqref{final}, that describes the optical response from an isolated high-quality resonant mode at any parameters of the incident plane wave, such as frequency, intensity, and the incident angle. The results are verified in comparison with straightforward FEM simulations. It is shown that the model accurately describes the effect of bistability which occurs under variation of the angle of incidence or the intensity of the incident wave. In particular, it is demonstrated that variation of the incident angle can optimize the coupling between the resonator and the incident waves leading to bistability with low intensity incident waves $W_0 = 0.35  {\rm \mu W/\mu m}^2$.
The obtained threshold value of intensity is two orders of magnitude smaller than that predicted for bistability threshold due to the Kerr effect in photonic crystal microcavities \cite{cowan2003optical} and of the same order as the thermo-optic threshold values for 
photonic crystal nanocavities \cite{haret2009extremely}. The latter, however, involves a more complicated design of stack or ladder cavities. The low intensity bistability is found to result in small temperature differences of a few degrees Kelvin. The temperature differences can be increased by operating away from the critical coupling or using a resonant mode with a smaller quality factor, see Fig.~\ref{Fig3}~(c).
It is found the bistability is extremely sensitive to the angle of incidence.  We speculate that the aforementioned effect paves a way for engineering micro/nano-devices allowing for a highly selective control of thermo-optic effect by variation of the angle of incidence. Finally, we point out that, unlike the real part of the dielectric permittivity, the imaginary part varies by several orders of magnitude in the visible to near infrared range \cite{Green2008, Schinke2015}. Thus, the bistablity threshold are extremely dependent on the resonant eigenfrequency and, consequently, on  the size of the resonator. The lowest bistability thresholds can be achieved by designing resonators with eigenfrequency tuned to a given material absorption rates, which meet the critical coupling condition.

We acknowledge financial support from state contract No FWES-2024-0003 of Kirensky Institute of Physics.


\begin{thebibliography}{47}%
\makeatletter
\providecommand \@ifxundefined [1]{%
 \@ifx{#1\undefined}
}%
\providecommand \@ifnum [1]{%
 \ifnum #1\expandafter \@firstoftwo
 \else \expandafter \@secondoftwo
 \fi
}%
\providecommand \@ifx [1]{%
 \ifx #1\expandafter \@firstoftwo
 \else \expandafter \@secondoftwo
 \fi
}%
\providecommand \natexlab [1]{#1}%
\providecommand \enquote  [1]{``#1''}%
\providecommand \bibnamefont  [1]{#1}%
\providecommand \bibfnamefont [1]{#1}%
\providecommand \citenamefont [1]{#1}%
\providecommand \href@noop [0]{\@secondoftwo}%
\providecommand \href [0]{\begingroup \@sanitize@url \@href}%
\providecommand \@href[1]{\@@startlink{#1}\@@href}%
\providecommand \@@href[1]{\endgroup#1\@@endlink}%
\providecommand \@sanitize@url [0]{\catcode `\\12\catcode `\$12\catcode
  `\&12\catcode `\#12\catcode `\^12\catcode `\_12\catcode `\%12\relax}%
\providecommand \@@startlink[1]{}%
\providecommand \@@endlink[0]{}%
\providecommand \url  [0]{\begingroup\@sanitize@url \@url }%
\providecommand \@url [1]{\endgroup\@href {#1}{\urlprefix }}%
\providecommand \urlprefix  [0]{URL }%
\providecommand \Eprint [0]{\href }%
\providecommand \doibase [0]{http://dx.doi.org/}%
\providecommand \selectlanguage [0]{\@gobble}%
\providecommand \bibinfo  [0]{\@secondoftwo}%
\providecommand \bibfield  [0]{\@secondoftwo}%
\providecommand \translation [1]{[#1]}%
\providecommand \BibitemOpen [0]{}%
\providecommand \bibitemStop [0]{}%
\providecommand \bibitemNoStop [0]{.\EOS\space}%
\providecommand \EOS [0]{\spacefactor3000\relax}%
\providecommand \BibitemShut  [1]{\csname bibitem#1\endcsname}%
\let\auto@bib@innerbib\@empty
\bibitem [{\citenamefont {Zograf}\ \emph {et~al.}(2017)\citenamefont {Zograf},
  \citenamefont {Petrov}, \citenamefont {Zuev}, \citenamefont {Dmitriev},
  \citenamefont {Milichko}, \citenamefont {Makarov},\ and\ \citenamefont
  {Belov}}]{zograf2017resonant}%
  \BibitemOpen
  \bibfield  {author} {\bibinfo {author} {\bibfnamefont {G.~P.}\ \bibnamefont
  {Zograf}}, \bibinfo {author} {\bibfnamefont {M.~I.}\ \bibnamefont {Petrov}},
  \bibinfo {author} {\bibfnamefont {D.~A.}\ \bibnamefont {Zuev}}, \bibinfo
  {author} {\bibfnamefont {P.~A.}\ \bibnamefont {Dmitriev}}, \bibinfo {author}
  {\bibfnamefont {V.~A.}\ \bibnamefont {Milichko}}, \bibinfo {author}
  {\bibfnamefont {S.~V.}\ \bibnamefont {Makarov}}, \ and\ \bibinfo {author}
  {\bibfnamefont {P.~A.}\ \bibnamefont {Belov}},\ }\href@noop {} {\bibfield
  {journal} {\bibinfo  {journal} {Nano letters}\ }\textbf {\bibinfo {volume}
  {17}},\ \bibinfo {pages} {2945} (\bibinfo {year} {2017})}\BibitemShut
  {NoStop}%
\bibitem [{\citenamefont {Khandekar}\ and\ \citenamefont
  {Rodriguez}(2017)}]{Khandekar17}%
  \BibitemOpen
  \bibfield  {author} {\bibinfo {author} {\bibfnamefont {C.}~\bibnamefont
  {Khandekar}}\ and\ \bibinfo {author} {\bibfnamefont {A.~W.}\ \bibnamefont
  {Rodriguez}},\ }\href@noop {} {\bibfield  {journal} {\bibinfo  {journal}
  {Applied Physics Letters}\ }\textbf {\bibinfo {volume} {111}},\ \bibinfo
  {pages} {083104} (\bibinfo {year} {2017})}\BibitemShut {NoStop}%
\bibitem [{\citenamefont {Aouassa}\ \emph {et~al.}(2017)\citenamefont
  {Aouassa}, \citenamefont {Mitsai}, \citenamefont {Syubaev}, \citenamefont
  {Pavlov}, \citenamefont {Zhizhchenko}, \citenamefont {Jadli}, \citenamefont
  {Hassayoun}, \citenamefont {Zograf}, \citenamefont {Makarov},\ and\
  \citenamefont {Kuchmizhak}}]{aouassa2017temperature}%
  \BibitemOpen
  \bibfield  {author} {\bibinfo {author} {\bibfnamefont {M.}~\bibnamefont
  {Aouassa}}, \bibinfo {author} {\bibfnamefont {E.}~\bibnamefont {Mitsai}},
  \bibinfo {author} {\bibfnamefont {S.}~\bibnamefont {Syubaev}}, \bibinfo
  {author} {\bibfnamefont {D.}~\bibnamefont {Pavlov}}, \bibinfo {author}
  {\bibfnamefont {A.}~\bibnamefont {Zhizhchenko}}, \bibinfo {author}
  {\bibfnamefont {I.}~\bibnamefont {Jadli}}, \bibinfo {author} {\bibfnamefont
  {L.}~\bibnamefont {Hassayoun}}, \bibinfo {author} {\bibfnamefont
  {G.}~\bibnamefont {Zograf}}, \bibinfo {author} {\bibfnamefont
  {S.}~\bibnamefont {Makarov}}, \ and\ \bibinfo {author} {\bibfnamefont
  {A.}~\bibnamefont {Kuchmizhak}},\ }\href@noop {} {\bibfield  {journal}
  {\bibinfo  {journal} {Applied Physics Letters}\ }\textbf {\bibinfo {volume}
  {111}} (\bibinfo {year} {2017})}\BibitemShut {NoStop}%
\bibitem [{\citenamefont {Celebrano}\ \emph {et~al.}(2021)\citenamefont
  {Celebrano}, \citenamefont {Rocco}, \citenamefont {Gandolfi}, \citenamefont
  {Zilli}, \citenamefont {Rusconi}, \citenamefont {Tognazzi}, \citenamefont
  {Mazzanti}, \citenamefont {Ghirardini}, \citenamefont {Pogna}, \citenamefont
  {Carletti}, \citenamefont {Baratto}, \citenamefont {Marino}, \citenamefont
  {Gigli}, \citenamefont {Biagioni}, \citenamefont {Du{\`{o}}}, \citenamefont
  {Cerullo}, \citenamefont {Leo}, \citenamefont {Valle}, \citenamefont
  {Finazzi},\ and\ \citenamefont {Angelis}}]{Celebrano21}%
  \BibitemOpen
  \bibfield  {author} {\bibinfo {author} {\bibfnamefont {M.}~\bibnamefont
  {Celebrano}}, \bibinfo {author} {\bibfnamefont {D.}~\bibnamefont {Rocco}},
  \bibinfo {author} {\bibfnamefont {M.}~\bibnamefont {Gandolfi}}, \bibinfo
  {author} {\bibfnamefont {A.}~\bibnamefont {Zilli}}, \bibinfo {author}
  {\bibfnamefont {F.}~\bibnamefont {Rusconi}}, \bibinfo {author} {\bibfnamefont
  {A.}~\bibnamefont {Tognazzi}}, \bibinfo {author} {\bibfnamefont
  {A.}~\bibnamefont {Mazzanti}}, \bibinfo {author} {\bibfnamefont
  {L.}~\bibnamefont {Ghirardini}}, \bibinfo {author} {\bibfnamefont {E.~A.~A.}\
  \bibnamefont {Pogna}}, \bibinfo {author} {\bibfnamefont {L.}~\bibnamefont
  {Carletti}}, \bibinfo {author} {\bibfnamefont {C.}~\bibnamefont {Baratto}},
  \bibinfo {author} {\bibfnamefont {G.}~\bibnamefont {Marino}}, \bibinfo
  {author} {\bibfnamefont {C.}~\bibnamefont {Gigli}}, \bibinfo {author}
  {\bibfnamefont {P.}~\bibnamefont {Biagioni}}, \bibinfo {author}
  {\bibfnamefont {L.}~\bibnamefont {Du{\`{o}}}}, \bibinfo {author}
  {\bibfnamefont {G.}~\bibnamefont {Cerullo}}, \bibinfo {author} {\bibfnamefont
  {G.}~\bibnamefont {Leo}}, \bibinfo {author} {\bibfnamefont {G.~D.}\
  \bibnamefont {Valle}}, \bibinfo {author} {\bibfnamefont {M.}~\bibnamefont
  {Finazzi}}, \ and\ \bibinfo {author} {\bibfnamefont {C.~D.}\ \bibnamefont
  {Angelis}},\ }\href {\doibase 10.1364/ol.420790} {\bibfield  {journal}
  {\bibinfo  {journal} {Optics Letters}\ }\textbf {\bibinfo {volume} {46}},\
  \bibinfo {pages} {2453} (\bibinfo {year} {2021})}\BibitemShut {NoStop}%
\bibitem [{\citenamefont {Cho}\ \emph {et~al.}(2023)\citenamefont {Cho},
  \citenamefont {Lee}, \citenamefont {Kim}, \citenamefont {Hu}, \citenamefont
  {Lee},\ and\ \citenamefont {Kim}}]{Cho2023}%
  \BibitemOpen
  \bibfield  {author} {\bibinfo {author} {\bibfnamefont {J.-W.}\ \bibnamefont
  {Cho}}, \bibinfo {author} {\bibfnamefont {Y.-J.}\ \bibnamefont {Lee}},
  \bibinfo {author} {\bibfnamefont {J.-H.}\ \bibnamefont {Kim}}, \bibinfo
  {author} {\bibfnamefont {R.}~\bibnamefont {Hu}}, \bibinfo {author}
  {\bibfnamefont {E.}~\bibnamefont {Lee}}, \ and\ \bibinfo {author}
  {\bibfnamefont {S.-K.}\ \bibnamefont {Kim}},\ }\href {\doibase
  10.1021/acsnano.3c01184} {\bibfield  {journal} {\bibinfo  {journal} {{ACS}
  Nano}\ }\textbf {\bibinfo {volume} {17}},\ \bibinfo {pages} {10442} (\bibinfo
  {year} {2023})}\BibitemShut {NoStop}%
\bibitem [{\citenamefont {Yang}\ \emph {et~al.}(2022)\citenamefont {Yang},
  \citenamefont {Chen}, \citenamefont {Zhao}, \citenamefont {Kim},
  \citenamefont {Luo},\ and\ \citenamefont {Hu}}]{Yang2022}%
  \BibitemOpen
  \bibfield  {author} {\bibinfo {author} {\bibfnamefont {F.}~\bibnamefont
  {Yang}}, \bibinfo {author} {\bibfnamefont {K.}~\bibnamefont {Chen}}, \bibinfo
  {author} {\bibfnamefont {Y.}~\bibnamefont {Zhao}}, \bibinfo {author}
  {\bibfnamefont {S.-K.}\ \bibnamefont {Kim}}, \bibinfo {author} {\bibfnamefont
  {X.}~\bibnamefont {Luo}}, \ and\ \bibinfo {author} {\bibfnamefont
  {R.}~\bibnamefont {Hu}},\ }\href {\doibase 10.1063/5.0076765} {\bibfield
  {journal} {\bibinfo  {journal} {Applied Physics Letters}\ }\textbf {\bibinfo
  {volume} {120}} (\bibinfo {year} {2022}),\ 10.1063/5.0076765}\BibitemShut
  {NoStop}%
\bibitem [{\citenamefont {Baffou}\ and\ \citenamefont
  {Quidant}(2013)}]{baffou2013thermo}%
  \BibitemOpen
  \bibfield  {author} {\bibinfo {author} {\bibfnamefont {G.}~\bibnamefont
  {Baffou}}\ and\ \bibinfo {author} {\bibfnamefont {R.}~\bibnamefont
  {Quidant}},\ }\href@noop {} {\bibfield  {journal} {\bibinfo  {journal} {Laser
  \& Photonics Reviews}\ }\textbf {\bibinfo {volume} {7}},\ \bibinfo {pages}
  {171} (\bibinfo {year} {2013})}\BibitemShut {NoStop}%
\bibitem [{\citenamefont {Danesi}\ \emph {et~al.}(2018)\citenamefont {Danesi},
  \citenamefont {Gandolfi}, \citenamefont {Carletti}, \citenamefont {Bontempi},
  \citenamefont {De~Angelis}, \citenamefont {Banfi},\ and\ \citenamefont
  {Alessandri}}]{danesi2018photo}%
  \BibitemOpen
  \bibfield  {author} {\bibinfo {author} {\bibfnamefont {S.}~\bibnamefont
  {Danesi}}, \bibinfo {author} {\bibfnamefont {M.}~\bibnamefont {Gandolfi}},
  \bibinfo {author} {\bibfnamefont {L.}~\bibnamefont {Carletti}}, \bibinfo
  {author} {\bibfnamefont {N.}~\bibnamefont {Bontempi}}, \bibinfo {author}
  {\bibfnamefont {C.}~\bibnamefont {De~Angelis}}, \bibinfo {author}
  {\bibfnamefont {F.}~\bibnamefont {Banfi}}, \ and\ \bibinfo {author}
  {\bibfnamefont {I.}~\bibnamefont {Alessandri}},\ }\href@noop {} {\bibfield
  {journal} {\bibinfo  {journal} {Physical Chemistry Chemical Physics}\
  }\textbf {\bibinfo {volume} {20}},\ \bibinfo {pages} {15307} (\bibinfo {year}
  {2018})}\BibitemShut {NoStop}%
\bibitem [{\citenamefont {Sun}\ and\ \citenamefont {Reano}(2010)}]{Sun10}%
  \BibitemOpen
  \bibfield  {author} {\bibinfo {author} {\bibfnamefont {P.}~\bibnamefont
  {Sun}}\ and\ \bibinfo {author} {\bibfnamefont {R.~M.}\ \bibnamefont
  {Reano}},\ }\href@noop {} {\bibfield  {journal} {\bibinfo  {journal} {Optics
  letters}\ }\textbf {\bibinfo {volume} {35}},\ \bibinfo {pages} {1124}
  (\bibinfo {year} {2010})}\BibitemShut {NoStop}%
\bibitem [{\citenamefont {Zograf}\ \emph {et~al.}(2018)\citenamefont {Zograf},
  \citenamefont {Yu}, \citenamefont {Baryshnikova}, \citenamefont {Kuznetsov},\
  and\ \citenamefont {Makarov}}]{zograf2018local}%
  \BibitemOpen
  \bibfield  {author} {\bibinfo {author} {\bibfnamefont {G.~P.}\ \bibnamefont
  {Zograf}}, \bibinfo {author} {\bibfnamefont {Y.~F.}\ \bibnamefont {Yu}},
  \bibinfo {author} {\bibfnamefont {K.~V.}\ \bibnamefont {Baryshnikova}},
  \bibinfo {author} {\bibfnamefont {A.~I.}\ \bibnamefont {Kuznetsov}}, \ and\
  \bibinfo {author} {\bibfnamefont {S.~V.}\ \bibnamefont {Makarov}},\
  }\href@noop {} {\bibfield  {journal} {\bibinfo  {journal} {JETP Letters}\
  }\textbf {\bibinfo {volume} {107}},\ \bibinfo {pages} {699} (\bibinfo {year}
  {2018})}\BibitemShut {NoStop}%
\bibitem [{\citenamefont {Pottier}\ and\ \citenamefont
  {Bellon}(2021)}]{Pottier21}%
  \BibitemOpen
  \bibfield  {author} {\bibinfo {author} {\bibfnamefont {B.}~\bibnamefont
  {Pottier}}\ and\ \bibinfo {author} {\bibfnamefont {L.}~\bibnamefont
  {Bellon}},\ }\href {\doibase 10.21468/scipostphys.10.5.120} {\bibfield
  {journal} {\bibinfo  {journal} {{SciPost} Physics}\ }\textbf {\bibinfo
  {volume} {10}},\ \bibinfo {pages} {120} (\bibinfo {year} {2021})}\BibitemShut
  {NoStop}%
\bibitem [{\citenamefont {Ryabov}\ \emph {et~al.}(2022)\citenamefont {Ryabov},
  \citenamefont {Pashina}, \citenamefont {Zograf}, \citenamefont {Makarov},\
  and\ \citenamefont {Petrov}}]{Ryabov22}%
  \BibitemOpen
  \bibfield  {author} {\bibinfo {author} {\bibfnamefont {D.}~\bibnamefont
  {Ryabov}}, \bibinfo {author} {\bibfnamefont {O.}~\bibnamefont {Pashina}},
  \bibinfo {author} {\bibfnamefont {G.}~\bibnamefont {Zograf}}, \bibinfo
  {author} {\bibfnamefont {S.}~\bibnamefont {Makarov}}, \ and\ \bibinfo
  {author} {\bibfnamefont {M.}~\bibnamefont {Petrov}},\ }\href {\doibase
  10.1515/nanoph-2022-0074} {\bibfield  {journal} {\bibinfo  {journal}
  {Nanophotonics}\ }\textbf {\bibinfo {volume} {11}},\ \bibinfo {pages} {3981}
  (\bibinfo {year} {2022})}\BibitemShut {NoStop}%
\bibitem [{\citenamefont {Gao}\ \emph {et~al.}(2017)\citenamefont {Gao},
  \citenamefont {Zhou}, \citenamefont {Sun}, \citenamefont {Tsang},\ and\
  \citenamefont {Shu}}]{Gao17a}%
  \BibitemOpen
  \bibfield  {author} {\bibinfo {author} {\bibfnamefont {Y.}~\bibnamefont
  {Gao}}, \bibinfo {author} {\bibfnamefont {W.}~\bibnamefont {Zhou}}, \bibinfo
  {author} {\bibfnamefont {X.}~\bibnamefont {Sun}}, \bibinfo {author}
  {\bibfnamefont {H.~K.}\ \bibnamefont {Tsang}}, \ and\ \bibinfo {author}
  {\bibfnamefont {C.}~\bibnamefont {Shu}},\ }\href@noop {} {\bibfield
  {journal} {\bibinfo  {journal} {Optics Letters}\ }\textbf {\bibinfo {volume}
  {42}},\ \bibinfo {pages} {1950} (\bibinfo {year} {2017})}\BibitemShut
  {NoStop}%
\bibitem [{\citenamefont {Jiang}\ and\ \citenamefont
  {Yang}(2020)}]{jiang2020optothermal}%
  \BibitemOpen
  \bibfield  {author} {\bibinfo {author} {\bibfnamefont {X.}~\bibnamefont
  {Jiang}}\ and\ \bibinfo {author} {\bibfnamefont {L.}~\bibnamefont {Yang}},\
  }\href@noop {} {\bibfield  {journal} {\bibinfo  {journal} {Light: Science \&
  Applications}\ }\textbf {\bibinfo {volume} {9}},\ \bibinfo {pages} {24}
  (\bibinfo {year} {2020})}\BibitemShut {NoStop}%
\bibitem [{\citenamefont {Tang}\ \emph {et~al.}(2021)\citenamefont {Tang},
  \citenamefont {Yen}, \citenamefont {Nishida}, \citenamefont {Takahara},
  \citenamefont {Zhang}, \citenamefont {Li}, \citenamefont {Fujita},\ and\
  \citenamefont {Chu}}]{tang2021mie}%
  \BibitemOpen
  \bibfield  {author} {\bibinfo {author} {\bibfnamefont {Y.-L.}\ \bibnamefont
  {Tang}}, \bibinfo {author} {\bibfnamefont {T.-H.}\ \bibnamefont {Yen}},
  \bibinfo {author} {\bibfnamefont {K.}~\bibnamefont {Nishida}}, \bibinfo
  {author} {\bibfnamefont {J.}~\bibnamefont {Takahara}}, \bibinfo {author}
  {\bibfnamefont {T.}~\bibnamefont {Zhang}}, \bibinfo {author} {\bibfnamefont
  {X.}~\bibnamefont {Li}}, \bibinfo {author} {\bibfnamefont {K.}~\bibnamefont
  {Fujita}}, \ and\ \bibinfo {author} {\bibfnamefont {S.-W.}\ \bibnamefont
  {Chu}},\ }\href@noop {} {\bibfield  {journal} {\bibinfo  {journal} {Optical
  Materials Express}\ }\textbf {\bibinfo {volume} {11}},\ \bibinfo {pages}
  {3608} (\bibinfo {year} {2021})}\BibitemShut {NoStop}%
\bibitem [{\citenamefont {Sivan}\ and\ \citenamefont
  {Chu}(2017)}]{sivan2017nonlinear}%
  \BibitemOpen
  \bibfield  {author} {\bibinfo {author} {\bibfnamefont {Y.}~\bibnamefont
  {Sivan}}\ and\ \bibinfo {author} {\bibfnamefont {S.-W.}\ \bibnamefont
  {Chu}},\ }\href@noop {} {\bibfield  {journal} {\bibinfo  {journal}
  {Nanophotonics}\ }\textbf {\bibinfo {volume} {6}},\ \bibinfo {pages} {317}
  (\bibinfo {year} {2017})}\BibitemShut {NoStop}%
\bibitem [{\citenamefont {Li}\ \emph {et~al.}(2021)\citenamefont {Li},
  \citenamefont {Tang}, \citenamefont {Takahara},\ and\ \citenamefont
  {Chu}}]{li2021nonlinear}%
  \BibitemOpen
  \bibfield  {author} {\bibinfo {author} {\bibfnamefont {C.-H.}\ \bibnamefont
  {Li}}, \bibinfo {author} {\bibfnamefont {Y.-L.}\ \bibnamefont {Tang}},
  \bibinfo {author} {\bibfnamefont {J.}~\bibnamefont {Takahara}}, \ and\
  \bibinfo {author} {\bibfnamefont {S.-W.}\ \bibnamefont {Chu}},\ }\href
  {\doibase 10.1063/5.0067251} {\bibfield  {journal} {\bibinfo  {journal} {The
  Journal of Chemical Physics}\ }\textbf {\bibinfo {volume} {155}} (\bibinfo
  {year} {2021}),\ 10.1063/5.0067251}\BibitemShut {NoStop}%
\bibitem [{\citenamefont {Duh}\ \emph {et~al.}(2020)\citenamefont {Duh},
  \citenamefont {Nagasaki}, \citenamefont {Tang}, \citenamefont {Wu},
  \citenamefont {Cheng}, \citenamefont {Yen}, \citenamefont {Ding},
  \citenamefont {Nishida}, \citenamefont {Hotta}, \citenamefont {Yang} \emph
  {et~al.}}]{duh2020giant}%
  \BibitemOpen
  \bibfield  {author} {\bibinfo {author} {\bibfnamefont {Y.-S.}\ \bibnamefont
  {Duh}}, \bibinfo {author} {\bibfnamefont {Y.}~\bibnamefont {Nagasaki}},
  \bibinfo {author} {\bibfnamefont {Y.-L.}\ \bibnamefont {Tang}}, \bibinfo
  {author} {\bibfnamefont {P.-H.}\ \bibnamefont {Wu}}, \bibinfo {author}
  {\bibfnamefont {H.-Y.}\ \bibnamefont {Cheng}}, \bibinfo {author}
  {\bibfnamefont {T.-H.}\ \bibnamefont {Yen}}, \bibinfo {author} {\bibfnamefont
  {H.-X.}\ \bibnamefont {Ding}}, \bibinfo {author} {\bibfnamefont
  {K.}~\bibnamefont {Nishida}}, \bibinfo {author} {\bibfnamefont
  {I.}~\bibnamefont {Hotta}}, \bibinfo {author} {\bibfnamefont {J.-H.}\
  \bibnamefont {Yang}},  \emph {et~al.},\ }\href@noop {} {\bibfield  {journal}
  {\bibinfo  {journal} {Nature communications}\ }\textbf {\bibinfo {volume}
  {11}},\ \bibinfo {pages} {4101} (\bibinfo {year} {2020})}\BibitemShut
  {NoStop}%
\bibitem [{\citenamefont {Zhang}\ \emph {et~al.}(2020)\citenamefont {Zhang},
  \citenamefont {Che}, \citenamefont {Chen}, \citenamefont {Xu}, \citenamefont
  {Xu}, \citenamefont {Wen}, \citenamefont {Lu}, \citenamefont {Liu},
  \citenamefont {Wang}, \citenamefont {Xu} \emph {et~al.}}]{zhang2020anapole}%
  \BibitemOpen
  \bibfield  {author} {\bibinfo {author} {\bibfnamefont {T.}~\bibnamefont
  {Zhang}}, \bibinfo {author} {\bibfnamefont {Y.}~\bibnamefont {Che}}, \bibinfo
  {author} {\bibfnamefont {K.}~\bibnamefont {Chen}}, \bibinfo {author}
  {\bibfnamefont {J.}~\bibnamefont {Xu}}, \bibinfo {author} {\bibfnamefont
  {Y.}~\bibnamefont {Xu}}, \bibinfo {author} {\bibfnamefont {T.}~\bibnamefont
  {Wen}}, \bibinfo {author} {\bibfnamefont {G.}~\bibnamefont {Lu}}, \bibinfo
  {author} {\bibfnamefont {X.}~\bibnamefont {Liu}}, \bibinfo {author}
  {\bibfnamefont {B.}~\bibnamefont {Wang}}, \bibinfo {author} {\bibfnamefont
  {X.}~\bibnamefont {Xu}},  \emph {et~al.},\ }\href@noop {} {\bibfield
  {journal} {\bibinfo  {journal} {Nature communications}\ }\textbf {\bibinfo
  {volume} {11}},\ \bibinfo {pages} {3027} (\bibinfo {year}
  {2020})}\BibitemShut {NoStop}%
\bibitem [{\citenamefont {Zograf}\ \emph {et~al.}(2021)\citenamefont {Zograf},
  \citenamefont {Petrov}, \citenamefont {Makarov},\ and\ \citenamefont
  {Kivshar}}]{zograf2021all}%
  \BibitemOpen
  \bibfield  {author} {\bibinfo {author} {\bibfnamefont {G.~P.}\ \bibnamefont
  {Zograf}}, \bibinfo {author} {\bibfnamefont {M.~I.}\ \bibnamefont {Petrov}},
  \bibinfo {author} {\bibfnamefont {S.~V.}\ \bibnamefont {Makarov}}, \ and\
  \bibinfo {author} {\bibfnamefont {Y.~S.}\ \bibnamefont {Kivshar}},\
  }\href@noop {} {\bibfield  {journal} {\bibinfo  {journal} {Advances in Optics
  and Photonics}\ }\textbf {\bibinfo {volume} {13}},\ \bibinfo {pages} {643}
  (\bibinfo {year} {2021})}\BibitemShut {NoStop}%
\bibitem [{\citenamefont {Kivshar}(2018)}]{kivshar2018all}%
  \BibitemOpen
  \bibfield  {author} {\bibinfo {author} {\bibfnamefont {Y.}~\bibnamefont
  {Kivshar}},\ }\href@noop {} {\bibfield  {journal} {\bibinfo  {journal}
  {National Science Review}\ }\textbf {\bibinfo {volume} {5}},\ \bibinfo
  {pages} {144} (\bibinfo {year} {2018})}\BibitemShut {NoStop}%
\bibitem [{\citenamefont {Baranov}\ \emph {et~al.}(2017)\citenamefont
  {Baranov}, \citenamefont {Zuev}, \citenamefont {Lepeshov}, \citenamefont
  {Kotov}, \citenamefont {Krasnok}, \citenamefont {Evlyukhin},\ and\
  \citenamefont {Chichkov}}]{baranov2017all}%
  \BibitemOpen
  \bibfield  {author} {\bibinfo {author} {\bibfnamefont {D.~G.}\ \bibnamefont
  {Baranov}}, \bibinfo {author} {\bibfnamefont {D.~A.}\ \bibnamefont {Zuev}},
  \bibinfo {author} {\bibfnamefont {S.~I.}\ \bibnamefont {Lepeshov}}, \bibinfo
  {author} {\bibfnamefont {O.~V.}\ \bibnamefont {Kotov}}, \bibinfo {author}
  {\bibfnamefont {A.~E.}\ \bibnamefont {Krasnok}}, \bibinfo {author}
  {\bibfnamefont {A.~B.}\ \bibnamefont {Evlyukhin}}, \ and\ \bibinfo {author}
  {\bibfnamefont {B.~N.}\ \bibnamefont {Chichkov}},\ }\href@noop {} {\bibfield
  {journal} {\bibinfo  {journal} {Optica}\ }\textbf {\bibinfo {volume} {4}},\
  \bibinfo {pages} {814} (\bibinfo {year} {2017})}\BibitemShut {NoStop}%
\bibitem [{\citenamefont {Saadabad}\ \emph {et~al.}(2021)\citenamefont
  {Saadabad}, \citenamefont {Huang},\ and\ \citenamefont
  {Miroshnichenko}}]{Saadabad21}%
  \BibitemOpen
  \bibfield  {author} {\bibinfo {author} {\bibfnamefont {R.~M.}\ \bibnamefont
  {Saadabad}}, \bibinfo {author} {\bibfnamefont {L.}~\bibnamefont {Huang}}, \
  and\ \bibinfo {author} {\bibfnamefont {A.~E.}\ \bibnamefont
  {Miroshnichenko}},\ }\href {\doibase 10.1103/physrevb.104.235405} {\bibfield
  {journal} {\bibinfo  {journal} {Physical Review B}\ }\textbf {\bibinfo
  {volume} {104}},\ \bibinfo {pages} {235405} (\bibinfo {year}
  {2021})}\BibitemShut {NoStop}%
\bibitem [{\citenamefont {Tan}\ \emph {et~al.}(2022)\citenamefont {Tan},
  \citenamefont {Yuan},\ and\ \citenamefont {Lu}}]{Tan22}%
  \BibitemOpen
  \bibfield  {author} {\bibinfo {author} {\bibfnamefont {L.}~\bibnamefont
  {Tan}}, \bibinfo {author} {\bibfnamefont {L.}~\bibnamefont {Yuan}}, \ and\
  \bibinfo {author} {\bibfnamefont {Y.~Y.}\ \bibnamefont {Lu}},\ }\href
  {\doibase 10.1364/josab.447417} {\bibfield  {journal} {\bibinfo  {journal}
  {Journal of the Optical Society of America B}\ }\textbf {\bibinfo {volume}
  {39}},\ \bibinfo {pages} {611} (\bibinfo {year} {2022})}\BibitemShut
  {NoStop}%
\bibitem [{\citenamefont {Maksimov}\ \emph {et~al.}(2022)\citenamefont
  {Maksimov}, \citenamefont {Kostyukov}, \citenamefont {Ershov}, \citenamefont
  {Molokeev}, \citenamefont {Bulgakov},\ and\ \citenamefont
  {Gerasimov}}]{Maksimov22}%
  \BibitemOpen
  \bibfield  {author} {\bibinfo {author} {\bibfnamefont {D.~N.}\ \bibnamefont
  {Maksimov}}, \bibinfo {author} {\bibfnamefont {A.~S.}\ \bibnamefont
  {Kostyukov}}, \bibinfo {author} {\bibfnamefont {A.~E.}\ \bibnamefont
  {Ershov}}, \bibinfo {author} {\bibfnamefont {M.~S.}\ \bibnamefont
  {Molokeev}}, \bibinfo {author} {\bibfnamefont {E.~N.}\ \bibnamefont
  {Bulgakov}}, \ and\ \bibinfo {author} {\bibfnamefont {V.~S.}\ \bibnamefont
  {Gerasimov}},\ }\href {\doibase 10.1103/PhysRevA.106.063507} {\bibfield
  {journal} {\bibinfo  {journal} {Physical Review A}\ }\textbf {\bibinfo
  {volume} {106}},\ \bibinfo {pages} {063507} (\bibinfo {year}
  {2022})}\BibitemShut {NoStop}%
\bibitem [{\citenamefont {Zhang}\ and\ \citenamefont
  {Zhang}(2015)}]{zhang2015ultrasensitive}%
  \BibitemOpen
  \bibfield  {author} {\bibinfo {author} {\bibfnamefont {M.}~\bibnamefont
  {Zhang}}\ and\ \bibinfo {author} {\bibfnamefont {X.}~\bibnamefont {Zhang}},\
  }\href@noop {} {\bibfield  {journal} {\bibinfo  {journal} {Scientific
  reports}\ }\textbf {\bibinfo {volume} {5}},\ \bibinfo {pages} {1} (\bibinfo
  {year} {2015})}\BibitemShut {NoStop}%
\bibitem [{\citenamefont {Wang}\ \emph {et~al.}(2020)\citenamefont {Wang},
  \citenamefont {Duan}, \citenamefont {Chen}, \citenamefont {Zhou},
  \citenamefont {Liu},\ and\ \citenamefont {Xiao}}]{wang2020controlling}%
  \BibitemOpen
  \bibfield  {author} {\bibinfo {author} {\bibfnamefont {X.}~\bibnamefont
  {Wang}}, \bibinfo {author} {\bibfnamefont {J.}~\bibnamefont {Duan}}, \bibinfo
  {author} {\bibfnamefont {W.}~\bibnamefont {Chen}}, \bibinfo {author}
  {\bibfnamefont {C.}~\bibnamefont {Zhou}}, \bibinfo {author} {\bibfnamefont
  {T.}~\bibnamefont {Liu}}, \ and\ \bibinfo {author} {\bibfnamefont
  {S.}~\bibnamefont {Xiao}},\ }\href@noop {} {\bibfield  {journal} {\bibinfo
  {journal} {Physical Review B}\ }\textbf {\bibinfo {volume} {102}},\ \bibinfo
  {pages} {155432} (\bibinfo {year} {2020})}\BibitemShut {NoStop}%
\bibitem [{\citenamefont {Sang}\ \emph {et~al.}(2021)\citenamefont {Sang},
  \citenamefont {Dereshgi}, \citenamefont {Hadibrata}, \citenamefont
  {Tanriover},\ and\ \citenamefont {Aydin}}]{sang2021highly}%
  \BibitemOpen
  \bibfield  {author} {\bibinfo {author} {\bibfnamefont {T.}~\bibnamefont
  {Sang}}, \bibinfo {author} {\bibfnamefont {S.~A.}\ \bibnamefont {Dereshgi}},
  \bibinfo {author} {\bibfnamefont {W.}~\bibnamefont {Hadibrata}}, \bibinfo
  {author} {\bibfnamefont {I.}~\bibnamefont {Tanriover}}, \ and\ \bibinfo
  {author} {\bibfnamefont {K.}~\bibnamefont {Aydin}},\ }\href@noop {}
  {\bibfield  {journal} {\bibinfo  {journal} {Nanomaterials}\ }\textbf
  {\bibinfo {volume} {11}},\ \bibinfo {pages} {484} (\bibinfo {year}
  {2021})}\BibitemShut {NoStop}%
\bibitem [{\citenamefont {Xiao}\ \emph {et~al.}(2021)\citenamefont {Xiao},
  \citenamefont {Wang}, \citenamefont {Duan}, \citenamefont {Liu},\ and\
  \citenamefont {Yu}}]{xiao2021engineering}%
  \BibitemOpen
  \bibfield  {author} {\bibinfo {author} {\bibfnamefont {S.}~\bibnamefont
  {Xiao}}, \bibinfo {author} {\bibfnamefont {X.}~\bibnamefont {Wang}}, \bibinfo
  {author} {\bibfnamefont {J.}~\bibnamefont {Duan}}, \bibinfo {author}
  {\bibfnamefont {T.}~\bibnamefont {Liu}}, \ and\ \bibinfo {author}
  {\bibfnamefont {T.}~\bibnamefont {Yu}},\ }\href@noop {} {\bibfield  {journal}
  {\bibinfo  {journal} {JOSA B}\ }\textbf {\bibinfo {volume} {38}},\ \bibinfo
  {pages} {1325} (\bibinfo {year} {2021})}\BibitemShut {NoStop}%
\bibitem [{\citenamefont {Cai}\ \emph {et~al.}(2022)\citenamefont {Cai},
  \citenamefont {Liu}, \citenamefont {Zhu}, \citenamefont {Wu},\ and\
  \citenamefont {Huang}}]{cai2022enhancing}%
  \BibitemOpen
  \bibfield  {author} {\bibinfo {author} {\bibfnamefont {Y.}~\bibnamefont
  {Cai}}, \bibinfo {author} {\bibfnamefont {X.}~\bibnamefont {Liu}}, \bibinfo
  {author} {\bibfnamefont {K.}~\bibnamefont {Zhu}}, \bibinfo {author}
  {\bibfnamefont {H.}~\bibnamefont {Wu}}, \ and\ \bibinfo {author}
  {\bibfnamefont {Y.}~\bibnamefont {Huang}},\ }\href@noop {} {\bibfield
  {journal} {\bibinfo  {journal} {Journal of Quantitative Spectroscopy and
  Radiative Transfer}\ }\textbf {\bibinfo {volume} {283}},\ \bibinfo {pages}
  {108150} (\bibinfo {year} {2022})}\BibitemShut {NoStop}%
\bibitem [{\citenamefont {Bikbaev}\ \emph {et~al.}(2021)\citenamefont
  {Bikbaev}, \citenamefont {Maksimov}, \citenamefont {Pankin}, \citenamefont
  {Chen},\ and\ \citenamefont {Timofeev}}]{Bikbaev21}%
  \BibitemOpen
  \bibfield  {author} {\bibinfo {author} {\bibfnamefont {R.~G.}\ \bibnamefont
  {Bikbaev}}, \bibinfo {author} {\bibfnamefont {D.~N.}\ \bibnamefont
  {Maksimov}}, \bibinfo {author} {\bibfnamefont {P.~S.}\ \bibnamefont
  {Pankin}}, \bibinfo {author} {\bibfnamefont {K.-P.}\ \bibnamefont {Chen}}, \
  and\ \bibinfo {author} {\bibfnamefont {I.~V.}\ \bibnamefont {Timofeev}},\
  }\href {\doibase 10.1364/oe.416132} {\bibfield  {journal} {\bibinfo
  {journal} {Optics Express}\ }\textbf {\bibinfo {volume} {29}},\ \bibinfo
  {pages} {4672} (\bibinfo {year} {2021})}\BibitemShut {NoStop}%
\bibitem [{\citenamefont {Villeneuve}\ \emph {et~al.}(1996)\citenamefont
  {Villeneuve}, \citenamefont {Fan},\ and\ \citenamefont
  {Joannopoulos}}]{villeneuve1996microcavities}%
  \BibitemOpen
  \bibfield  {author} {\bibinfo {author} {\bibfnamefont {P.~R.}\ \bibnamefont
  {Villeneuve}}, \bibinfo {author} {\bibfnamefont {S.}~\bibnamefont {Fan}}, \
  and\ \bibinfo {author} {\bibfnamefont {J.}~\bibnamefont {Joannopoulos}},\
  }\href@noop {} {\bibfield  {journal} {\bibinfo  {journal} {Physical Review
  B}\ }\textbf {\bibinfo {volume} {54}},\ \bibinfo {pages} {7837} (\bibinfo
  {year} {1996})}\BibitemShut {NoStop}%
\bibitem [{\citenamefont {Painter}\ \emph {et~al.}(1999)\citenamefont
  {Painter}, \citenamefont {Vu{\v{c}}kovi{\'c}},\ and\ \citenamefont
  {Scherer}}]{painter1999defect}%
  \BibitemOpen
  \bibfield  {author} {\bibinfo {author} {\bibfnamefont {O.}~\bibnamefont
  {Painter}}, \bibinfo {author} {\bibfnamefont {J.}~\bibnamefont
  {Vu{\v{c}}kovi{\'c}}}, \ and\ \bibinfo {author} {\bibfnamefont
  {A.}~\bibnamefont {Scherer}},\ }\href@noop {} {\bibfield  {journal} {\bibinfo
   {journal} {JOSA B}\ }\textbf {\bibinfo {volume} {16}},\ \bibinfo {pages}
  {275} (\bibinfo {year} {1999})}\BibitemShut {NoStop}%
\bibitem [{\citenamefont {Hsu}\ \emph {et~al.}(2016)\citenamefont {Hsu},
  \citenamefont {Zhen}, \citenamefont {Stone}, \citenamefont {Joannopoulos},\
  and\ \citenamefont {Solja{\v{c}}i{\'c}}}]{Hsu16}%
  \BibitemOpen
  \bibfield  {author} {\bibinfo {author} {\bibfnamefont {C.~W.}\ \bibnamefont
  {Hsu}}, \bibinfo {author} {\bibfnamefont {B.}~\bibnamefont {Zhen}}, \bibinfo
  {author} {\bibfnamefont {A.~D.}\ \bibnamefont {Stone}}, \bibinfo {author}
  {\bibfnamefont {J.~D.}\ \bibnamefont {Joannopoulos}}, \ and\ \bibinfo
  {author} {\bibfnamefont {M.}~\bibnamefont {Solja{\v{c}}i{\'c}}},\ }\href
  {\doibase 10.1038/natrevmats.2016.48} {\bibfield  {journal} {\bibinfo
  {journal} {Nature Reviews Materials}\ }\textbf {\bibinfo {volume} {1}},\
  \bibinfo {pages} {16048} (\bibinfo {year} {2016})}\BibitemShut {NoStop}%
\bibitem [{\citenamefont {Koshelev}\ \emph {et~al.}(2019)\citenamefont
  {Koshelev}, \citenamefont {Favraud}, \citenamefont {Bogdanov}, \citenamefont
  {Kivshar},\ and\ \citenamefont {Fratalocchi}}]{Koshelev19}%
  \BibitemOpen
  \bibfield  {author} {\bibinfo {author} {\bibfnamefont {K.}~\bibnamefont
  {Koshelev}}, \bibinfo {author} {\bibfnamefont {G.}~\bibnamefont {Favraud}},
  \bibinfo {author} {\bibfnamefont {A.}~\bibnamefont {Bogdanov}}, \bibinfo
  {author} {\bibfnamefont {Y.}~\bibnamefont {Kivshar}}, \ and\ \bibinfo
  {author} {\bibfnamefont {A.}~\bibnamefont {Fratalocchi}},\ }\href {\doibase
  10.1515/nanoph-2019-0024} {\bibfield  {journal} {\bibinfo  {journal}
  {Nanophotonics}\ }\textbf {\bibinfo {volume} {8}},\ \bibinfo {pages} {725}
  (\bibinfo {year} {2019})}\BibitemShut {NoStop}%
\bibitem [{\citenamefont {Sadreev}(2021)}]{sadreev2021interference}%
  \BibitemOpen
  \bibfield  {author} {\bibinfo {author} {\bibfnamefont {A.~F.}\ \bibnamefont
  {Sadreev}},\ }\href@noop {} {\bibfield  {journal} {\bibinfo  {journal}
  {Reports on Progress in Physics}\ } (\bibinfo {year} {2021})}\BibitemShut
  {NoStop}%
\bibitem [{\citenamefont {Rybin}\ \emph {et~al.}(2017)\citenamefont {Rybin},
  \citenamefont {Koshelev}, \citenamefont {Sadrieva}, \citenamefont {Samusev},
  \citenamefont {Bogdanov}, \citenamefont {Limonov},\ and\ \citenamefont
  {Kivshar}}]{rybin2017high}%
  \BibitemOpen
  \bibfield  {author} {\bibinfo {author} {\bibfnamefont {M.~V.}\ \bibnamefont
  {Rybin}}, \bibinfo {author} {\bibfnamefont {K.~L.}\ \bibnamefont {Koshelev}},
  \bibinfo {author} {\bibfnamefont {Z.~F.}\ \bibnamefont {Sadrieva}}, \bibinfo
  {author} {\bibfnamefont {K.~B.}\ \bibnamefont {Samusev}}, \bibinfo {author}
  {\bibfnamefont {A.~A.}\ \bibnamefont {Bogdanov}}, \bibinfo {author}
  {\bibfnamefont {M.~F.}\ \bibnamefont {Limonov}}, \ and\ \bibinfo {author}
  {\bibfnamefont {Y.~S.}\ \bibnamefont {Kivshar}},\ }\href@noop {} {\bibfield
  {journal} {\bibinfo  {journal} {Physical review letters}\ }\textbf {\bibinfo
  {volume} {119}},\ \bibinfo {pages} {243901} (\bibinfo {year}
  {2017})}\BibitemShut {NoStop}%
\bibitem [{\citenamefont {Huang}\ \emph {et~al.}(2021)\citenamefont {Huang},
  \citenamefont {Xu}, \citenamefont {Rahmani}, \citenamefont {Neshev},\ and\
  \citenamefont {Miroshnichenko}}]{huang2021pushing}%
  \BibitemOpen
  \bibfield  {author} {\bibinfo {author} {\bibfnamefont {L.}~\bibnamefont
  {Huang}}, \bibinfo {author} {\bibfnamefont {L.}~\bibnamefont {Xu}}, \bibinfo
  {author} {\bibfnamefont {M.}~\bibnamefont {Rahmani}}, \bibinfo {author}
  {\bibfnamefont {D.}~\bibnamefont {Neshev}}, \ and\ \bibinfo {author}
  {\bibfnamefont {A.~E.}\ \bibnamefont {Miroshnichenko}},\ }\href@noop {}
  {\bibfield  {journal} {\bibinfo  {journal} {Advanced Photonics}\ }\textbf
  {\bibinfo {volume} {3}},\ \bibinfo {pages} {016004} (\bibinfo {year}
  {2021})}\BibitemShut {NoStop}%
\bibitem [{\citenamefont {Shadrina}\ \emph {et~al.}(2023)\citenamefont
  {Shadrina}, \citenamefont {Bulgakov}, \citenamefont {Maksimov},\ and\
  \citenamefont {Gerasimov}}]{shadrina2023thermo}%
  \BibitemOpen
  \bibfield  {author} {\bibinfo {author} {\bibfnamefont {G.}~\bibnamefont
  {Shadrina}}, \bibinfo {author} {\bibfnamefont {E.}~\bibnamefont {Bulgakov}},
  \bibinfo {author} {\bibfnamefont {D.}~\bibnamefont {Maksimov}}, \ and\
  \bibinfo {author} {\bibfnamefont {V.}~\bibnamefont {Gerasimov}},\ }\href@noop
  {} {\bibfield  {journal} {\bibinfo  {journal} {Physical Review B}\ }\textbf
  {\bibinfo {volume} {108}},\ \bibinfo {pages} {155411} (\bibinfo {year}
  {2023})}\BibitemShut {NoStop}%
\bibitem [{\citenamefont {Barulin}\ \emph {et~al.}(2024)\citenamefont
  {Barulin}, \citenamefont {Pashina}, \citenamefont {Riabov}, \citenamefont
  {Sergaeva}, \citenamefont {Sadrieva}, \citenamefont {Shcherbakov},
  \citenamefont {Rutckaia}, \citenamefont {Schilling}, \citenamefont
  {Bogdanov}, \citenamefont {Sinev} \emph {et~al.}}]{barulin2024thermo}%
  \BibitemOpen
  \bibfield  {author} {\bibinfo {author} {\bibfnamefont {A.}~\bibnamefont
  {Barulin}}, \bibinfo {author} {\bibfnamefont {O.}~\bibnamefont {Pashina}},
  \bibinfo {author} {\bibfnamefont {D.}~\bibnamefont {Riabov}}, \bibinfo
  {author} {\bibfnamefont {O.}~\bibnamefont {Sergaeva}}, \bibinfo {author}
  {\bibfnamefont {Z.}~\bibnamefont {Sadrieva}}, \bibinfo {author}
  {\bibfnamefont {A.}~\bibnamefont {Shcherbakov}}, \bibinfo {author}
  {\bibfnamefont {V.}~\bibnamefont {Rutckaia}}, \bibinfo {author}
  {\bibfnamefont {J.}~\bibnamefont {Schilling}}, \bibinfo {author}
  {\bibfnamefont {A.}~\bibnamefont {Bogdanov}}, \bibinfo {author}
  {\bibfnamefont {I.}~\bibnamefont {Sinev}},  \emph {et~al.},\ }\href@noop {}
  {\bibfield  {journal} {\bibinfo  {journal} {Laser \& Photonics Reviews}\
  }\textbf {\bibinfo {volume} {18}},\ \bibinfo {pages} {2301399} (\bibinfo
  {year} {2024})}\BibitemShut {NoStop}%
\bibitem [{\citenamefont {Fan}\ \emph {et~al.}(2003)\citenamefont {Fan},
  \citenamefont {Suh},\ and\ \citenamefont {Joannopoulos}}]{Fan03}%
  \BibitemOpen
  \bibfield  {author} {\bibinfo {author} {\bibfnamefont {S.}~\bibnamefont
  {Fan}}, \bibinfo {author} {\bibfnamefont {W.}~\bibnamefont {Suh}}, \ and\
  \bibinfo {author} {\bibfnamefont {J.~D.}\ \bibnamefont {Joannopoulos}},\
  }\href {\doibase 10.1364/josaa.20.000569} {\bibfield  {journal} {\bibinfo
  {journal} {Journal of the Optical Society of America A}\ }\textbf {\bibinfo
  {volume} {20}},\ \bibinfo {pages} {569} (\bibinfo {year} {2003})}\BibitemShut
  {NoStop}%
\bibitem [{\citenamefont {Ruan}\ and\ \citenamefont {Fan}(2012)}]{Ruan12}%
  \BibitemOpen
  \bibfield  {author} {\bibinfo {author} {\bibfnamefont {Z.}~\bibnamefont
  {Ruan}}\ and\ \bibinfo {author} {\bibfnamefont {S.}~\bibnamefont {Fan}},\
  }\href {\doibase 10.1103/physreva.85.043828} {\bibfield  {journal} {\bibinfo
  {journal} {Physical Review A}\ }\textbf {\bibinfo {volume} {85}} (\bibinfo
  {year} {2012}),\ 10.1103/physreva.85.043828}\BibitemShut {NoStop}%
\bibitem [{\citenamefont {Christopoulos}\ \emph {et~al.}(2019)\citenamefont
  {Christopoulos}, \citenamefont {Tsilipakos}, \citenamefont {Sinatkas},\ and\
  \citenamefont {Kriezis}}]{christopoulos2019calculation}%
  \BibitemOpen
  \bibfield  {author} {\bibinfo {author} {\bibfnamefont {T.}~\bibnamefont
  {Christopoulos}}, \bibinfo {author} {\bibfnamefont {O.}~\bibnamefont
  {Tsilipakos}}, \bibinfo {author} {\bibfnamefont {G.}~\bibnamefont
  {Sinatkas}}, \ and\ \bibinfo {author} {\bibfnamefont {E.~E.}\ \bibnamefont
  {Kriezis}},\ }\href@noop {} {\bibfield  {journal} {\bibinfo  {journal}
  {Optics Express}\ }\textbf {\bibinfo {volume} {27}},\ \bibinfo {pages}
  {14505} (\bibinfo {year} {2019})}\BibitemShut {NoStop}%
\bibitem [{\citenamefont {Cowan}\ and\ \citenamefont
  {Young}(2003)}]{cowan2003optical}%
  \BibitemOpen
  \bibfield  {author} {\bibinfo {author} {\bibfnamefont {A.~R.}\ \bibnamefont
  {Cowan}}\ and\ \bibinfo {author} {\bibfnamefont {J.~F.}\ \bibnamefont
  {Young}},\ }\href@noop {} {\bibfield  {journal} {\bibinfo  {journal}
  {Physical Review E}\ }\textbf {\bibinfo {volume} {68}},\ \bibinfo {pages}
  {046606} (\bibinfo {year} {2003})}\BibitemShut {NoStop}%
\bibitem [{\citenamefont {Haret}\ \emph {et~al.}(2009)\citenamefont {Haret},
  \citenamefont {Tanabe}, \citenamefont {Kuramochi},\ and\ \citenamefont
  {Notomi}}]{haret2009extremely}%
  \BibitemOpen
  \bibfield  {author} {\bibinfo {author} {\bibfnamefont {L.-D.}\ \bibnamefont
  {Haret}}, \bibinfo {author} {\bibfnamefont {T.}~\bibnamefont {Tanabe}},
  \bibinfo {author} {\bibfnamefont {E.}~\bibnamefont {Kuramochi}}, \ and\
  \bibinfo {author} {\bibfnamefont {M.}~\bibnamefont {Notomi}},\ }\href@noop {}
  {\bibfield  {journal} {\bibinfo  {journal} {Optics express}\ }\textbf
  {\bibinfo {volume} {17}},\ \bibinfo {pages} {21108} (\bibinfo {year}
  {2009})}\BibitemShut {NoStop}%
\bibitem [{\citenamefont {Green}(2008)}]{Green2008}%
  \BibitemOpen
  \bibfield  {author} {\bibinfo {author} {\bibfnamefont {M.~A.}\ \bibnamefont
  {Green}},\ }\href {\doibase 10.1016/j.solmat.2008.06.009} {\bibfield
  {journal} {\bibinfo  {journal} {Solar Energy Materials and Solar Cells}\
  }\textbf {\bibinfo {volume} {92}},\ \bibinfo {pages} {1305} (\bibinfo {year}
  {2008})}\BibitemShut {NoStop}%
\bibitem [{\citenamefont {Schinke}\ \emph {et~al.}(2015)\citenamefont
  {Schinke}, \citenamefont {Christian~Peest}, \citenamefont {Schmidt},
  \citenamefont {Brendel}, \citenamefont {Bothe}, \citenamefont {Vogt},
  \citenamefont {Kr\"oger}, \citenamefont {Winter}, \citenamefont
  {Schirmacher}, \citenamefont {Lim}, \citenamefont {Nguyen},\ and\
  \citenamefont {MacDonald}}]{Schinke2015}%
  \BibitemOpen
  \bibfield  {author} {\bibinfo {author} {\bibfnamefont {C.}~\bibnamefont
  {Schinke}}, \bibinfo {author} {\bibfnamefont {P.}~\bibnamefont
  {Christian~Peest}}, \bibinfo {author} {\bibfnamefont {J.}~\bibnamefont
  {Schmidt}}, \bibinfo {author} {\bibfnamefont {R.}~\bibnamefont {Brendel}},
  \bibinfo {author} {\bibfnamefont {K.}~\bibnamefont {Bothe}}, \bibinfo
  {author} {\bibfnamefont {M.~R.}\ \bibnamefont {Vogt}}, \bibinfo {author}
  {\bibfnamefont {I.}~\bibnamefont {Kr\"oger}}, \bibinfo {author}
  {\bibfnamefont {S.}~\bibnamefont {Winter}}, \bibinfo {author} {\bibfnamefont
  {A.}~\bibnamefont {Schirmacher}}, \bibinfo {author} {\bibfnamefont
  {S.}~\bibnamefont {Lim}}, \bibinfo {author} {\bibfnamefont {H.~T.}\
  \bibnamefont {Nguyen}}, \ and\ \bibinfo {author} {\bibfnamefont
  {D.}~\bibnamefont {MacDonald}},\ }\href {\doibase 10.1063/1.4923379}
  {\bibfield  {journal} {\bibinfo  {journal} {AIP Advances}\ }\textbf {\bibinfo
  {volume} {5}} (\bibinfo {year} {2015}),\ 10.1063/1.4923379}\BibitemShut
  {NoStop}%
\end{thebibliography}

%

\end{document}